# 環境適応処理における運用開始後FPGAロジック変更の提案

山登庸次†

† NTT ネットワークサービスシステム研究所，東京都武蔵野市緑町 3-9-11
E-mail: †yoji.yamato.wa@hco.ntt.co.jp

**あらまし** 近年，ヘテロジニアスハードウェア利用が増えているが，これらの十分な活用には，OpenCL 等のハードウェアを意識した技術理解が必要であり，壁は高いのが現状である．これらの背景から，私は，エンジニアが通常 CPU 向けに記述したアプリケーションコードを，配置される環境に応じて，自動で変換やリソース量設定等をして，高性能に運用可能とする環境適応ソフトウェアのコンセプトを提案してきた．本稿では，今まで運用開始前の変換や設定しか検討していなかったが，運用中の利用特性に応じて，すでに動作している FPGA ロジックを再構成することを検証する．運用中 FPGA 再構成方式を提案し，FPGA に自動オフロードしたアプリケーションを，運用中再構成により別アプリケーションに FPGA ロジックを再構成する処理を通じて，性能改善と断時間を確認し，有効性を示す．
**キーワード** 環境適応ソフトウェア, 自動オフロード，FPGA，運用中再構成，コストパフォーマンス

# Proposal of FPGA logic change after service launch for environment adapation

Yoji YAMATO†

† Network Service Systems Laboratories, NTT Corporation, 3-9-11, Midori-cho, Musashino-shi, Tokyo
E-mail: †yoji.yamato.wa@hco.ntt.co.jp

**Abstract** In order to make full use of heterogeneous hardware, it is necessary to have a technical skill of hardware such as OpenCL, and the current situation is that the barrier is high. Based on this background, I have proposed environment-adaptive software that enables high-performance operation by automatically converting application code written for normal CPUs by engineers according to the deployed environment and setting appropriate amount of resources. Until now, I only considered conversions and settings before the start of operation. In this paper, I verify that the logic is reconfigured according to the usage characteristics during operation. I confirm that the application running on the FPGA is reconfigured into another application according to the usage characteristics.
**Key words** Environment Adaptive Software, Automatic Offloading, FPGA, Reconfiguration during Operation, Cost Performance

## 1. はじめに

ムーアの法則の減速予測から，CPU（Central Processing Unit）1 コアの半導体集積度やクロック数を上げるだけでなく，コア数を増やすマルチコア CPU，GPU（Graphics Processing Unit）や FPGA（Field Programmable Gate Array）等のヘテロジニアスなハードウェアが，通常のアプリケーション運用に用いられるようになっている．Microsoft 社は FPGA の検索利用等の取り組みをしているし [1]，Amazon 社はクラウドを用いて（例えば, [2]- [7]），FPGA や GPU インスタンスを提供している [8]．また，ヘテロジニアスなハードウェアとして，IoT 機器等の小型デバイスの利用（ [9]- [18]）も増えている．

しかし，1 コアの CPU でないヘテロジニアスなハードウェアを効率よく利用するためには，ハードウェアに応じたプログラム作成や設定が必要となり，大半の技術者にとっては，壁が高い．マルチコア CPU では OpenMP（Open Multi-Processing）[19]，GPU では CUDA（Compute Unified Device Architecture）[20]，FPGA では OpenCL（Open Computing Language）[21]，IoT 機器ではアセンブリ等の高度な知識が必要となってくることが多い．

ヘテロジニアスなハードウェアをより活用していくためには，高度な知識を持たない通常の技術者でも，それらを最大限活用できるようにするプラットフォームが必要と考える．技術者が通常の CPU と同様のロジックで処理を記述したソフトウェアを，分析して，配置先の環境（マルチコア CPU，GPU，FPGA 等）にあわせて，適切に変換，設定を行い，環境に適応した動作をさせることを，プラットフォームが行うことが今後求められていく．

そこで，私は，一度記述したプログラムコードを，配置先の環境に存在する GPU や FPGA，マルチコア CPU 等を利用できるように，変換，リソース設定，配置決定等を自動で行い，アプリケーションを高性能に動作させることを目的とした，環境適応ソフトウェアを提案している．合わせて，環境適応ソフトウェアの要素として，コードのループ文及び機能ブロックを，GPU，FPGA に自動オフロードする方式，GPU 等の処理リソース量を適切にアサインする方式等を提案して評価している [22]- [27].

しかし，私の環境適応はこれまで，アプリケーションの運用開始前に変換や配置等の適応処理を行うことが前提となっており，運用開始後に利用特性変化等に応じて再構成することは想定されていなかった．本稿は，運用中に利用特性に応じて再構成する中で，特に FPGA ロジックの再構成を対象とする．まず，通常の CPU 向けプログラムを FPGA にオフロードして運用開始し，リクエスト特性を分析し，別プログラムに FPGA ロジックを変更することを提案し，ユーザ影響少なく再構成する手法を検討，評価する．検討手法の有効性を，既存アプリケーションの FPGA 構成と運用中再構成を通じて，確認する．

## 2. 既存技術

### 2.1 市中技術

GPU の単純な計算力を一般的計算にも用いる GPGPU（General Purpose GPU）[28] が近年盛んになってきており，そのための環境として NVIDIA は CUDA を提供している．GPU だけに限定せずに，FPGA，GPU 等のヘテロジニアスなハードウェアを共通的に扱う仕様としては OpenCL があり，多くのベンダが OpenCL に対応してきている．OpenCL や CUDA は，C 言語の拡張を用いてプログラムを記述する．拡張記述として，カーネルと呼ばれる FPGA 等とホストと呼ばれる CPU の間のメモリ情報の移行などを記述するが，オリジナルの C 言語に比べてハードウェアの知識が必要と言われる．

OpenCL や CUDA の文法を理解していなくても，容易に GPU 等のヘテロジニアスなハードウェアを用いることができるようにするため，ディレクティブで GPU 処理等を行う行を指定して，ディレクティブに基づいてコンパイラが，GPU やマルチコア CPU のバイナリファイルを作成する取り組みがある．OpenMP や OpenACC [29] 等の仕様が，それを解釈実行する gcc や PGI [30] 等のコンパイラがある．

整理すると，OpenMP や OpenACC 等を用いることで容易に，また，OpenCL や CUDA 等を用いることでより細かく，FPGA や GPU，マルチコア CPU を利用できるようになっている．しかし，それらのハードウェアを利用することはできても，性能改善は容易ではないのが現状である．例えば，Intel コンパイラ [31] という，自動で CPU の複数のコアに処理を分配するコンパイラがある．Intel コンパイラ等は，自動化時は，プログラムのループの中で，並列処理可能なループを見つけ，複数のコアに処理を行わせている．しかし，データコピー等により，単に複数のコアでループを処理しても性能が改善しないことも多い．マルチコア CPU でなく，GPU や FPGA の際は

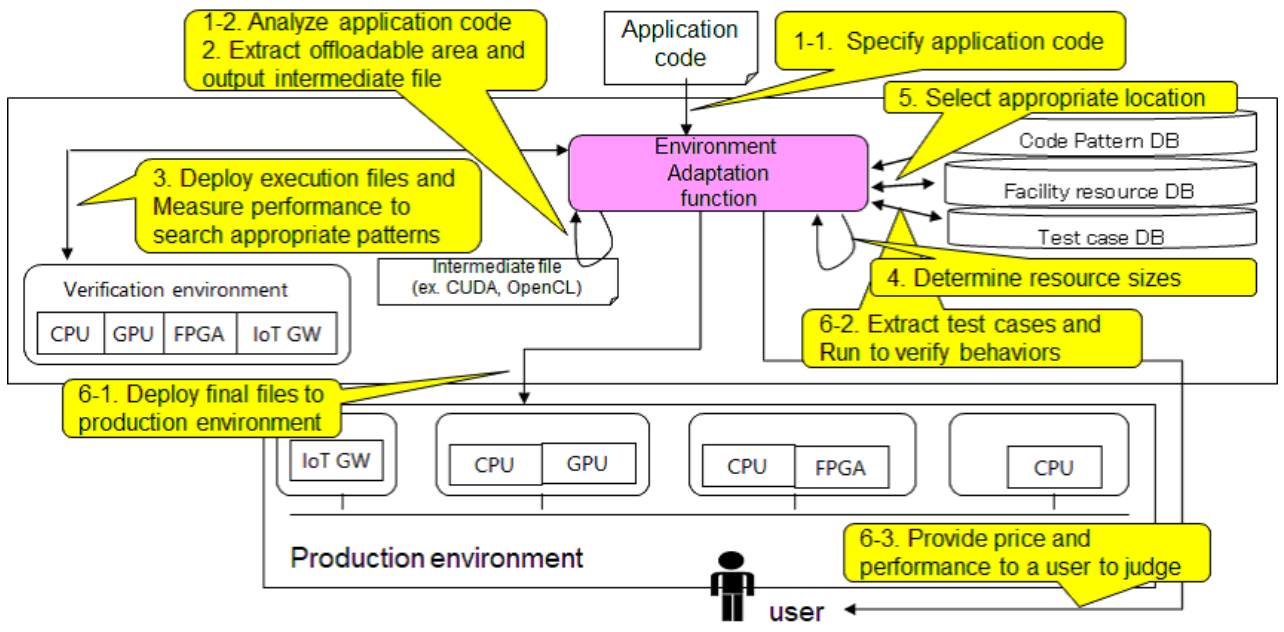

図 1　環境適応ソフトウェアの全体像

メモリも異なるためより複雑で，性能改善には，OpenCL や CUDA を駆使してチューニングが必要だったり，gcc 等を用いて適切な GPU 処理箇所を試行錯誤で探索することが必要だったりする．このようにヘテロジニアスなハードウェアを用いた性能改善には，技術スキルが必要だったり，試行錯誤の稼働が必要だったりする．

ループ文の GPU オフロードとして，ループ文の GPU 処理箇所探索を自動化する取り組みとして，私は進化計算手法である遺伝的アルゴリズム（GA）[32] を用いたオフロードを提案している．また，FPGA では，コンパイルに長時間かかり，何度も測定することができないため，ループ文算術強度やオフロード時 FPGA リソース使用率を元に，候補とするループ文を絞った後に OpenCL 化して測定して適切なパターンを探索する手法を提案している．

### 2.2 環境適応ソフトウェア

以前に私は，環境適応ソフトウェアの処理として，図 1 の処理全体像を提案した．環境適応ソフトウェアの処理は，事業者が提供する環境適応機能を中心として，商用環境，検証環境，コードパターン DB，設備リソース DB，テストケース DB 等がプラットフォーム機能として連携して行われる．

ここで，図の Step1-6 は，アプリケーションの運用開始前に必要となる，コードの変換，リソース量の調整，配置場所の調整，検証を行うが，Step7 は，アプリケーションの運用開始後に，利用特性等を分析して，必要な再構成を行う．再構成の対象は，運用開始前と同様に，コード変換，リソース量の調整，配置場所の調整である．

課題を整理する．ヘテロジニアスなハードウェアに対するオフロードは手動での取組みが主流である．私は環境適応ソフトウェアのコンセプトを提案し，ループ文等の GPU や FPGA 自動オフロード方式を検証しているが全て運用開始前の話で，Step7 に当たる運用開始後の再構成については検討がされていない．そのため，本稿では，運用中再構成の中で，FPGA オフロードロジックの，利用特性に応じた再構成の方式を対象とする．再構成の対象は，GPU，FPGA オフロードロジック，リソース量，配置場所等，多々あるが，利用特性に応じて FPGA ロジックを再構成する例は，商用クラウドでも例がなく，インパクトが大きいため，FPGA ロジックを対象とする．

## 3. 運用中 FPGA 再構成

### 3.1 運用開始前 FPGA 自動オフロード方式のレビュー

著者以前論文で検証しているループ文の FPGA 自動オフロード方式を図 2 でレビューする [27].

本手法は，まず，オフロードしたいソースコードの分析を Clang [33] 等で行い，ループ文や変数の情報を把握する．

次に，把握したループ文に対して，FPGA オフロードを試行するかどうか候補を絞っていく．ループ文がオフロード効果があるかどうかは，算術強度が一つの指標となりうる．算術強度は，計算回数が多いと増加し，データサイズが大きいと減少する指標で，算術強度が高い処理は重い処理となり時間がかかる．そこで，ループ文の算術強度を分析し，強度が高いループ文をオフロード候補に絞る．算術強度分析には ROSE framework [34] を用いた．また，ループ回数が多いループも重い処理となる．ループ回数はプロファイラーで分析し，ループ回数が多いループ文もオフロード候補に絞る．ループ回数分析には gcov を用いた．

高算術強度やループ回数が多いループ文であっても，それを FPGA で処理する際に，リソースを過度に消費してしまうのは問題である．そこで次に，ループ文を FPGA 処理する際のリソース量の算出に移る．FPGA にコンパイルする際，OpenCL 等の高位言語からハードウェア記述の HDL（Hardware Decription Language）等のレベルに変換され，それに基づき配線処理等がされる．この時，配線処理等は多大な時間がかかるが，HDL 等の途中状態の段階までは時間は分単位でしかかからない．HDL レベルで，FPGA 利用リソースは分かるため，利用リソース量は短時間で分かる．オフロード候補のループ文を OpenCL 言語化し，リソース量を算出することで，オフロードした際の算術強度とリソース量が決まるため，算術強度/リソース量をリソース効率とする．本手法では，高リソース効率のループ文をオフロード候補として更に絞り込む．ここで，ループ文を OpenCL 言語化する際には，CPU 処理のプログラムを，カーネル（FPGA）とホスト（CPU）に，OpenCL の文法に従って分割する．

次に，高リソース効率のループ文が幾つか絞られるため，それらを用いて性能測定するパターンを作成する．絞り込まれた単ループ文とその組み合わせのパターンを一定数作り，FPGA で動作するようコンパイルする．最後に検証環境で，コンパイルされた複数パターンの性能測定を行い，高速のパターンを解として選択する．

### 3.2 運用中 FPGA 再構成に向けた基本方針

3.1 節の手法で，ユーザが指定したアプリケーションで，FPGA に適したループ文部分を FPGA に自動オフロードすることができる．ユーザが使う商用環境にオフロード後，商用環境での実際の性能と価格を確認し，ユーザはアプリケーションを利用開始する．ただし，3.1 節でのオフロードで用いる，性能最適化用テストケース（複数のオフロードパターンで性能比較する際に性能測定する項目）は，ユーザが指定した，運用開始前の想定利用データを利用しており，運用開始後に実利用されるデータから大きく離れる可能性がある．

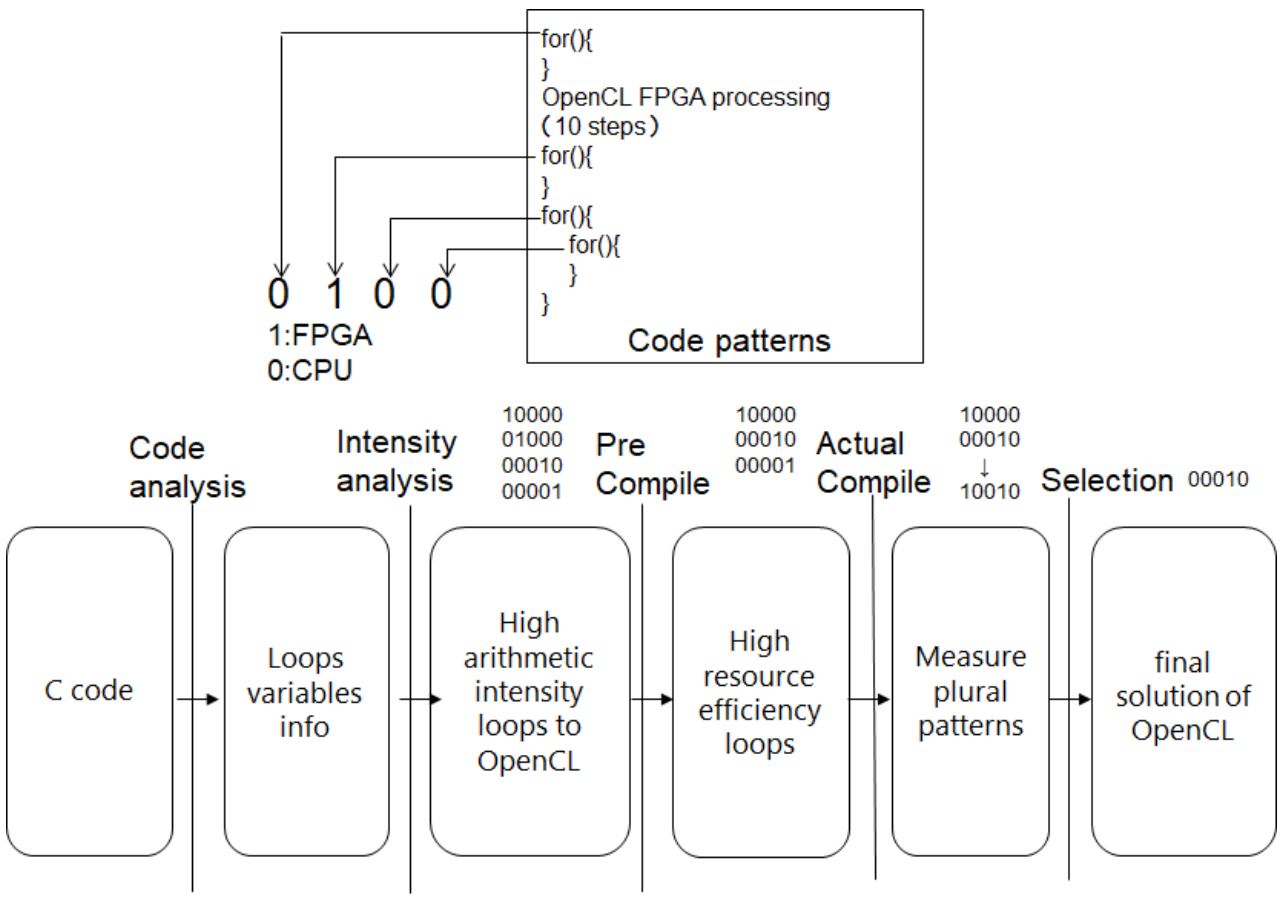

図 2 ループ文の FPGA 自動オフロード方式

そのため，3.2 節では，運用開始後の利用形態が，最初の想定と異なり，FPGA には別ロジックをオフロードした方が性能が向上する等の場合に，FPGA ロジックをユーザ影響低く再構成することを検討する．再構成は，同じアプリケーションでも異なるループ文オフロードに変える場合もあれば，異なるアプリケーションのオフロードに変える場合もある．

FPGA の再構成は，動的再構成，静的再構成の 2 つがあり，前者は FPGA を動かしながら回路構成を変更する技術で，書換のための断時間は ms のオーダー，後者は FPGA を停止してから回路構成を変更する技術で，断時間は 1 秒程度である．断時間のユーザ影響度によって，FPGA を製造するベンダの提供する再構成手法を選択すればよいが，どちらでも断時間は発生することや，別ロジックへの書換は動作確認の試験が必要なことから，頻繁に再構成するべきとは考えておらず，効果が閾値以上の場合だけ提案する等制限を設ける．

検討する再構成は，1 か月等，一定期間のリクエスト傾向の分析から始まる．リクエスト傾向を分析し，現在オフロードしているアプリケーションより処理負荷が高いか同等のものがあるかを把握する．次に，処理負荷が高いリクエストを，想定利用データでなく実際に商用で利用されているデータを使って，FPGA オフロードの最適化試行を検証環境で行う．ここで，検証により見つかった新しいオフロードパターンが現在のオフロードパターンより十分改善効果が高いかを閾値以上かそれ以下かで判定する．閾値を上回る場合は，ユーザに再構成を提案する．ユーザ了承後，商用環境を再構成するが，できるだけユーザ影響を抑えて再構成する．

### 3.3 運用中 FPGA 再構成方式の提案

3.2 節の基本方針に基づいて，3.3 節では具体的な運用中再構成方式を提案する．再構成方式は，6 ステップからなり，各ステップを詳細説明する．特にステップ 1 は複雑であるため，最後に補足説明する．

1. 一定期間（長期間）の，商用リクエストデータ履歴を分析し，処理時間負荷上位の複数アプリケーションを特定し，そのアプリケーション利用時の代表データを取得する．

1-1. 一定期間の各アプリケーション利用履歴から，実処理時間と利用回数合計を計算．ただし，FPGA オフロードされているアプリケーションでは，オフロードされなかった場合の処理時間を仮に計算．運用開始前の想定利用データでの試験履歴から，(CPU 処理のみの際の実処理時間)/(FPGA オフロードされた際の実処理時間) で改善度係数を求めておく．次に，実処理時間に改善度係数をかけた値の合計を比較に用いる処理時間合計とする．

1-2. 実処理時間合計を全アプリケーションで比較．

1-3. 実処理時間合計順に並べかえ，処理時間負荷上位の複数アプリケーションを特定．

1-4. 負荷上位アプリケーションの一定期間（短期間）のリクエストデータを取得し，データサイズを一定サイズごとに整列させ度数分布を作成．

1-5. データサイズ度数分布の最頻値 Mode に該当する実リクエストデータから，どれか一つデータを選び，代表データに選定．

2. 複数の負荷上位アプリケーションで，商用代表データのテストケースを高速化する，オフロードパターンを検証環境測定を通じて抽出．

2-1. 負荷上位アプリケーションで，算術強度が高い 4 つの for 文を選択．

2-2. 4for 文をオフロードする 4OpenCL を作成し，プレコンパイルして，リソース使用率を求め，算術強度/リソース使用率が高い 3 つの for 文を選択．

2-3. 3OpenCL を代表データで性能測定する．性能上位 2 つの for 文を組合せた OpenCL を作成し同様に性能測定する．

2-4. 4 測定で最高速のオフロードパターンを解とする．

3. 現オフロードパターンと抽出した複数の新オフロードパターンの処理時間を測定し，商用利用頻度に基づく性能改善効果を求める．商用代表データでのテストケースで，

3-1. 現オフロードパターンで (検証環境実処理時間削減)*(商用環境利用頻度) 計算．

3-2. 複数の新オフロードパターンで (検証環境実処理時間削減)*(商用環境利用頻度) 計算．

4. 新オフロードパターンの性能改善度効果が現オフロードパターンのそれの閾値以上であるかで，再構成提案を判断．

4-1. 複数の (3-2) ÷ (3-1) を計算し，閾値以上かを確認し，以上なら再構成提案，以下なら何もしない．

5. 契約ユーザに FPGA 再構成実施を提案し，OK/NG の返答を得る．

6. 商用環境で別 OpenCL を起動することで静的再構成を実施．

6-1. 新オフロードパターンのコンパイル．

6-2. 現オフロードパターンの動作を停止．

6-3. 新オフロードパターンの動作を起動．

ステップ 1 で負荷上位アプリケーションを選定するが，FPGA オフロードされているアプリケーションについては，改善度係数をかけることで，オフロードされなかった場合を計算し，CPU 処理のみに補正して比較する．また，代表データを選ぶ際は，データサイズの平均では実利用データと大きく異なる場合もあるので，データサイズの最頻値 Mode を使う．

## 4. 評　価

### 4.1 評価条件

#### 4.1.1 評価対象

評価対象は，多くのユーザが FPGA で利用すると想定される信号処理と画像処理を中心にする．

信号処理の有限インパルス応答フィルタ（tdFIR）は，システムにインパルス関数を入力したときの出力に対して有限時間で打ち切る処理を行う，フィルタの一種である．実装は種々あるが, [35] の C コードを用いる．IoT 等で，デバイスからの信号データをネットワーク転送するアプリケーションを考えた際に，ネットワークコストを下げるため，フィルタ等の信号処理をしてからクラウドにデータを送ることは想定される．そのため，信号処理の FPGA での自動高速化は応用範囲が広いと考える．

MRI-Q [36] は，キャリブレーションのためのスキャナー設定を表現する Q マトリックを計算する MRI 画像処理である．MRI-Q は非カルテアン空間で 3D MRI 再構成アルゴリズムで使用される．IoT 等で，画像処理はしばしばカメラ映像の自動監視等に必要となり，画像処理のスループット等の性能は，強化が要望される．オフロードパターン抽出時の性能測定では，MRI-Q は 3D MRI 画像処理を実行し，データサイズによるが，想定利用では，64*64*64 サイズのデータを使用して処理時間を測定する．

この他は，非圧縮流体解析の姫野ベンチマーク [37]，対称行列計算の Symm [38]，離散フーリエ変換計算の DFT (Discrete Fourier Transform) [39] を，同じサーバ上で動かし，実行リクエストを受けるとする．

#### 4.1.2 評価手法

運用開始前は，tdFIR のオフロードをユーザが指定し，自動オフロードする．商用環境には，tdFIR だけ FPGA オフロードし，MRI-Q，姫野ベンチマーク，Symm，DFT は CPU のみの処理で動作させる．一定期間，商用環境サーバにリクエスト負荷をかけて，その結果を分析し，性能改善効果が高いオフロードパターンへの再構成を提案し，ユーザ承認後再構成されることを確認する．

FPGA オフロード時の条件は以下で行う．

オフロード対象：ループ文数　tdFIR 6，MRI-Q 16，姫野ベンチマーク 13，Symm 9，DFT 10.

算術強度絞り込み：算術強度分析の上位 4 つのループ文に絞り込み

リソース効率絞り込み：リソース効率分析の上位 3 つのループ文に絞り込み

実測オフロードパターン数：4（1 回目は上位 3 つのループ文オフロードパターンを測定し，2 回目は 1 回目で高性能だった 2 つのループ文オフロードの組合せパターンを測定.）

FPGA 再構成に向けた，運用の条件は以下で行う．

リクエスト負荷：tdFIR 300req/h，MRI-Q 10req/h，姫野

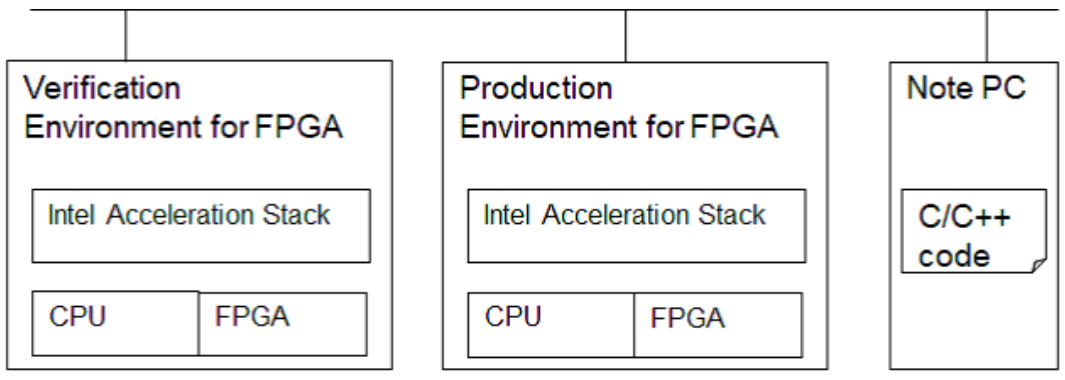

| Name | Hardware | CPU | RAM | FPGA | OS | Intel Acceleration Stack |
|---|---|---|---|---|---|---|
| Verification Environment for FPGA | Dell PowerEdge R740 | Intel(R) Xeon Bronze 3206R *2 | 32GB *4 | Intel PAC D5005 (Intel Stratix 10 GX FPGA) | CentOS 7.9 | 2.0 |
| Production Environment for FPGA | Dell PowerEdge R740 | Intel(R) Xeon Bronze 3206R *2 | 32GB *4 | Intel PAC D5005 (Intel Stratix 10 GX FPGA) | CentOS 7.9 | 2.0 |
| Note PC | HP ProBook 470 G3 | Intel Core i5-6200U @2.3GHz | 8GB | | Windows 10 Pro | |

図 3　性能測定環境

3req/h, Symm 2req/h, DFT 1req/h を 1 時間リクエストする．データはサンプルデータについている，Small，Large と Large を 1 回コピー追加して 2 倍にした 3 種を用意する．tdFIR, MRI-Q では，サイズ小中大のリクエストを 3:5:2 の比でリクエストする．姫野，Symm，DFT はサンプルデータのままとする．

負荷分析時の長期間：1 時間

代表データ選定時の短期間：1 時間

負荷上位アプリケーションの数：2

性能改善効果閾値：2.0

再構成実施の中で，再構成に伴う，性能改善効果と各ステップの処理時間を取得する．

#### 4.1.3　評 価 環 境

評価用 FPGA として Intel FPGA PAC D5005 (Intel Stratix 10 GX FPGA，LE 2,800,000) を用いる．Intel FPGA PAC D5005 搭載サーバは，DELL EMC PowerEdge R740 (CPU：Intel Xeon Bronze 3206R *2，RAM：32GB RDIMM * 4) である．FPGA の制御は，Intel Acceleration Stack Version 2.0 を用いる．OpenCL の文法に従い，C 言語プログラムを，カーネルプログラムとホストプログラムに分割記載することで，FPGA オフロード処理が OpenCL でされ，別 OpenCL プログラムへの再構成も，Intel Acceleration Stack で処理される．

評価環境を図 3 に示す．ここで，ノート PC が，オフロードするアプリケーションコードを指定し，検証環境での性能測定を通じてオフロードパターンを抽出後，商用環境にデプロイされる．商用環境アプリケーション群には，ノート PC から定期的にリクエストがされる．商用環境履歴を分析して，検証環境を用いて新たなオフロードパターンを抽出し，ユーザ確認後，商用環境を新オフロードパターンに再構成する．

### 4.2　結　　果

図 4 は，再構成前と再構成後のオフロードアプリケーションの処理時間改善度とそれに関連する一定期間の処理時間合計（改善度係数補正済み）を示している．まず，再構成前は tdFIR がオフロードされており，運用開始前の想定データでの改善度は 2.07 であり，運用開始後は 300req/h の負荷がかかっている．リクエストの実処理時間合計*2.07 の 79.7 秒が補正済処理時間合計で，300 が利用回数合計である．次に，MRI-Q は，運用開始後は 10req/h の負荷がかかっている．リクエストの実処理時間合計の 274 秒が処理時間合計で，10 が利用回数合計である．この 2 つが負荷上位のアプリケーションとなる．この 2 つで代表データを使って新オフロードパターンを探索する．新オフロードパターンでの 1 回の処理時間は，tdFIR は 0.266 秒から 0.129 秒に削減され，MRI-Q は 27.4 秒から 2.23 秒に削減されており，商用利用回数をかけることで，処理時間削減の性能改善度は，tdFIR は 41.1sec/h で，MRI-Q は 252sec/h となる．

| | Application | Improvement of processing time | Summation of processing time |
|---|---|---|---|
| Before reconfiguratoin | tdFIR | 41.1 sec/h | 79.7 sec |
| After reconfiguratoin | MRI-Q | 252 sec/h | 274 sec |

図 4　提案方式の再構成を通じた処理時間改善比較

図 4 から，tdFIR オフロードから，MRI-Q へのオフロードへの変更で，性能改善度は 6.1 倍となり，2.0 を超えるため，再構成がユーザに提案される．

リクエスト分析は今回 1 時間分のデータしか行っていないためサイズは小さいが，サイズに比例して長時間化すると考える．今回，リクエスト分析と代表データ選定は 1 秒程度，改善効果計算 1 日，再構成実施 1 秒程度がかかっている．運用開始前のオフロード試行や，運用中の新オフロードパターン試行，商用環境へ新オフロードパターンをコンパイルする時間については，1 回の FPGA コンパイルが 6 時間以上かかることから，測定数 4 つの場合は 1 アプリケーションでも 1 日以上かかる．新オフロードパターン試行を含め，分析等殆どの処理は，商用環境のアプリケーション運用中のバックグラウンドで行われるため，ユーザ影響はない．唯一，商用での再構成実施については，アプリケーションの断時間が発生するが，OpenCL の静的再構成でも 1 秒程度であり，殆ど影響がないことを確認できた．もし，より，短い断時間が必要な際は，Intel や Xilinx の動的再構成機能の利用も考えられる．

運用中の利用特性に応じた FPGA 再構成を，運用開始前の tdFIR オフロードから，運用中の MRI-Q オフロードへの変更で確認した．再構成を通じて，性能改善度が閾値以上に上がり，断時間が十分短く再構成できることを示した．

## 5.　ま と め

本稿では，アプリケーション運用開始後に，利用特性に応じて，適切な FPGA ロジックに運用中に再構成する，運用中 FPGA 再構成方式を提案した．

運用開始前に，アプリケーションループ文を FPGA に自動オフロードさせておく．提案方式は，一定期間毎に，実リクエストデータから，CPU 処理時間の負荷が上位のアプリケーションを取得し，該当する代表テストケースを把握する．次に，複数の代表テストケースを高速化するオフロードパターンを検証環境での試行測定を通じ抽出する．これは，運用開始前オフロードとほぼ同様である．次に，現在の高速化パターンと抽出した新しい高速化パターンの処理時間を測定し，商用利用頻度に基づく処理時間改善を求める．ここで，新しい高速化パター

ンが現在パターンの閾値以上効果の場合，再構成実施をユーザに提案する．ユーザ了承が得られたら，商用環境で OpenCL 再構成を用いて，変更を行う．FPGA に自動オフロードしたアプリケーションを，運用中再構成により別アプリケーションに FPGA ロジックを再構成した．再構成により処理時間削減を改善し，併せて，1 秒程度の短い断時間で再構成を行い，方式有効性を確認した．

## 文　献